\documentstyle{aipproc}
\begin{document}
\title{On Quantization of Field Theories in 
Polymomentum Variables\thanks{Submitted July 1998. To 
appear in: {\em Particles, Fields and Gravitation,} Proc. Int. Conf.,  
Lodz 1998, eds. K. Smolinski and J. Rembielinski (AIP Proc., 1998).}
}  
\author{Igor V. Kanatchikov$^*$
}  
\address{$^*$ Laboratory of Analytical Mechanics and Field Theory \\
Institute of Fundamental Technological Research \\
Polish Academy of Sciences \\ 
\'Swi\c etokrzyska 21, Warsaw PL-00-049, Poland \\ 
{ \footnotesize \tt e-mail: ikanat@ippt.gov.pl} }
 
\maketitle  

\vspace*{-74mm}
\hbox to 5.4truein{
\vspace*{-1mm}\footnotesize 
\hfil 
\hbox to 0 truecm{ 
\hss \normalsize hep-th/9811016} 
}
\vspace*{70mm}

\begin{abstract}
Polymomentum canonical theories, 
which are manifestly covariant 
multi-parameter generalizations 
of the Hamiltonian formalism  to field theory,  
are 
considered 
as a possible basis of quantization.  
We arrive at a  multi-parameter hypercomplex 
generalization of 
quantum mechanics to field theory 
in which 
the algebra of 
complex numbers and a time parameter 
are replaced 
by 
the space-time Clifford algebra 
and space-time variables  
treated in a manifestly covariant fashion. 
The corresponding  covariant   
generalization of 
the Schr\"odinger equation 
is shown to be consistent 
with several aspects of the correspondence 
principle such as  a relation to 
the 
De Donder-Weyl 
Hamilton-Jacobi theory  in the classical limit 
and the Ehrenfest theorem. 
A relation of 
the corresponding wave function 
(over 
a finite dimensional configuration space 
of field and space-time variables)      
with    the Schr\"odinger wave functional in quantum field theory 
is examined 
     in  the ultra-local approximation. 
\end{abstract} 

\newcommand{\sometext}{
The Hamiltonian formalism is a basis of 
most of the quantization procedures, the stochastic quantization 
and the path integral quantization being the 
remarkable exceptions. 
The conventional Hamiltonian formalism in field theory requires a 
singling out of a time parameter. This implies the global 
hyperbolicity of the underlying space-time manifold. 
When applying this framework 
to the problem of quantization of General Relativity 
a  concern  about its  applicability arises.   
In fact, already qualitative considerations demonstrate that on 
quantum level the space-time undergoes intricate  fluctuations of 
the metric and topology leading to the so-called space-time foam, 
for which  
no global hyperbolicity can be expected. 
}


\newcommand{\beq}{\begin{equation}}
\newcommand{\eeq}{\end{equation}}
\newcommand{\beqa}{\begin{eqnarray}}
\newcommand{\eeqa}{\end{eqnarray}}
\newcommand{\nn}{\nonumber}

\newcommand{\half}{\frac{1}{2}}

\newcommand{\xt}{\tilde{X}}

\newcommand{\uind}[2]{^{#1_1 \, ... \, #1_{#2}} }
\newcommand{\lind}[2]{_{#1_1 \, ... \, #1_{#2}} }
\newcommand{\com}[2]{[#1,#2]_{-}} 
\newcommand{\acom}[2]{[#1,#2]_{+}} 
\newcommand{\compm}[2]{[#1,#2]_{\pm}}

\newcommand{\lie}[1]{\pounds_{#1}}
\newcommand{\co}{\circ}
\newcommand{\sgn}[1]{(-1)^{#1}}
\newcommand{\lbr}[2]{ [ \hspace*{-1.5pt} [ #1 , #2 ] \hspace*{-1.5pt} ] }
\newcommand{\lbrpm}[2]{ [ \hspace*{-1.5pt} [ #1 , #2 ] \hspace*{-1.5pt}
 ]_{\pm} }
\newcommand{\lbrp}[2]{ [ \hspace*{-1.5pt} [ #1 , #2 ] \hspace*{-1.5pt} ]_+ }
\newcommand{\lbrm}[2]{ [ \hspace*{-1.5pt} [ #1 , #2 ] \hspace*{-1.5pt} ]_- }
\newcommand{\pbr}[2]{ \{ \hspace*{-2.2pt} [ #1 , #2 ] \hspace*{-2.55pt} \} }
\newcommand{\we}{\wedge}
\newcommand{\dv}{d^V}
\newcommand{\nbrpq}[2]{\nbr{\xxi{#1}{1}}{\xxi{#2}{2}}}
\newcommand{\lieni}[2]{$\pounds$${}_{\stackrel{#1}{X}_{#2}}$  }

\newcommand{\rbox}[2]{\raisebox{#1}{#2}}
\newcommand{\xx}[1]{\raisebox{1pt}{$\stackrel{#1}{X}$}}
\newcommand{\xxi}[2]{\raisebox{1pt}{$\stackrel{#1}{X}$$_{#2}$}}
\newcommand{\ff}[1]{\raisebox{1pt}{$\stackrel{#1}{F}$}}
\newcommand{\dd}[1]{\raisebox{1pt}{$\stackrel{#1}{D}$}}
\newcommand{\nbr}[2]{{\bf[}#1 , #2{\bf ]}}
\newcommand{\der}{\partial}
\newcommand{\oo}{$\Omega$}
\newcommand{\Om}{\Omega}
\newcommand{\om}{\omega}
\newcommand{\eps}{\epsilon}
\newcommand{\si}{\sigma}
\newcommand{\Lm}{\bigwedge^*}

\newcommand{\inn}{\hspace*{2pt}\raisebox{-1pt}{\rule{6pt}{.3pt}\hspace*
{0pt}\rule{.3pt}{8pt}\hspace*{3pt}}}
\newcommand{\sro}{Schr\"{o}dinger\ }
\newcommand{\bm}{\boldmath}
\newcommand{\vol}{\omega}

\newcommand{\dvol}[1]{\der_{#1}\inn \vol} 

\newcommand{\bd}{\mbox{\bf d}}
\newcommand{\bder}{\mbox{\bm $\der$}}
\newcommand{\bI}{\mbox{\bm $I$}}

\newcommand{\de}{\delta} 
 \newcommand{\be}{\beta}
\newcommand{\ga}{\gamma} 
\newcommand{\Ga}{\Gamma} 
\newcommand{\gmu}{\gamma^\mu}
\newcommand{\gnu}{\gamma^\nu}
\newcommand{\ka}{\kappa}
\newcommand{\hka}{\hbar \kappa}
\newcommand{\al}{\alpha}
\newcommand{\lapl}{\bigtriangleup}
\newcommand{\psib}{\overline{\psi}}
\newcommand{\Psib}{\overline{\Psi}}
\newcommand{\derts}{\stackrel{\leftrightarrow}{\der}}
\newcommand{\what}[1]{\widehat{#1}}

\newcommand{\bx}{{\bf x}}
\newcommand{\bk}{{\bf k}}
\newcommand{\bq}{{\bf q}}

\newcommand{\omk}{\omega_{\bf k}}

\def\ZZ{\mbox{\rm Z}\hskip-5pt \mbox{\rm Z}}
\def\RR{\mbox{\rm I}\hskip-2pt \mbox{\rm R}}
\def\CC{\mbox{\rm C}\hskip-5.5pt \mbox{l} \;}

\newcommand{\Eop}{\what{\mbox{$\cal E$}}}



\section*{Introduction }

The canonical quantization is based on the Hamiltonian 
formalism. 
The conventional Hamiltonian formalism in field theory is 
 an infinite dimensional version of the one in mechanics. 
As a result, 
the  
quantum field theory based on it is 
essentially 
the 
quantum mechanics of systems with an infinite number 
of degrees of freedom. 
Most of the difficulties and ambiguities 
of quantum field theory are due to this infinite dimensionality.  
However, should  quantum fields always be understood in this 
way? Does this picture exhaust all aspects of quantum fields? 
Is there a  
``genuine quantum field theory'' 
more general  
that just 
quantum mechanics applied to fields?   
It is clear that in 
pertubative regime, i.e. in 
the 
vicinity 
of a free field theory 
which can be 
represented as a continuum of 
harmonic oscillators,   the above picture can work well, and it 
really does as 
the 
experimental triumph of pertubative 
quantum field theory  demonstrates. 
However, applicability of this picture 
in non-pertubative domain 
and in curved space-time, where no natural particle concept exists 
in general, 
\nopagebreak[2]can   
be more limited. 

A  
conceivable 
approach to the above posed  questions can be 
based on the (not  widely 
acknowledged yet)  
fact that 
the conventional 
version of 
the Hamiltonian formalism in field theory is not 
the only one possible. 
 In fact, there exist  different alternative extensions 
of the Hamiltonian  formulation 
to field theory   
which all reduce 
to the Hamilton  formalism in mechanics 
if the number of space-time  dimensions equals to one. 
These extensions originate from the calculus of variations 
of multiple integrals \cite{dw,rund,giaquinta,kastrup}.   
Unlike the conventional Hamiltonian formalism, 
all these formulations are constructed in a manifestly covariant way 
not requiring any singling out of a time dimension. They can 
be applied even if the signature of the space-time is not 
Minkowskian. 
This is achieved by assigning the canonical momentum like variables, 
which we called {\em polymomenta} \cite{ikanat0},   
to the whole set of space-time 
derivatives of a field:   
$\der_\mu y^a \rightarrow p^\mu_a$\footnote{Throughout the paper 
$y^a$ denote 
field variables, $x^\mu$ are space-time variables 
($\mu=1,...,n$),  
$\der_\mu y^a$ are 
space-time derivatives (or first jets) of field variables,  
$p_a^\mu$ denote polymomenta.}.   
An analogue of the phase space is then a  
finite dimensional phase space of variables 
$(y^a, p_a^\mu, x^\nu)$  
which we call  the {\em polymomentum phase space}.  
Corresponding 
generalizations of the canonical formalism will be referred to 
as  {\em polymomentum canonical theories}. 
In the geometric (Cartan's) approach to the calculus of variations  
these theories 
(a version of which is also known 
as the multisymplectic formalism \cite{gimm}) 
appear as 
a 
result of a certain choice of the so-called  Lepagean 
equivalents of 
a 
field-theoretic (multidimensional) analogue 
of the Poincar\'e-Cartan form \cite{kastrup,lepage,gimm,gotay,dedecker}.    
\newcommand{\ignore}{  
Some mathematical issues related to this kind of 
theories, also known as the multisymplectic formalism \cite{gimm},   
have been studied recently 
\cite{math} 
from various points of view and 
using a language different from that we adopted in the present paper. 
}
Unfortunately, applications of these theories in 
physics have been so far rather rare (see for references 
\cite{kastrup,ikanat0,gimm}). 

The simplest example of a  polymomentum canonical theory 
is 
the so-called  De Donder-Weyl (DW) 
theory\cite{dw,rund,kastrup,gimm}.  
Given a Lagrangian density 
$L=L(y^a, \der_\mu y^a, x^\nu)$, 
the polymomenta are introduced by 
the formula $p_a^\mu:=\der L/ \der (\der_\mu y^a)$. An analogue of the 
Hamilton canonical function defined as  
$H:= \der_\mu y^a p_a^\mu - L$ is referred to as 
the {\em DW Hamiltonian function} in what follows. 
Note, that $H$  is a function 
of variables $(y^a, p_a^\mu,x^\mu)=:z^M$.  
In  these variables the Euler-Lagrange field 
equations can be rewritten in 
the form of {\em DW Hamiltonian field equations}  
\beq
\der_\mu y^a = \der H / \der p^\mu_a, 
\quad \der_\mu p^\mu_a = - \der H/ \der y^a . 
\eeq
Clearly, this formulation reproduces the standard Hamiltonian formulation 
in mechanics at $n=1$. At $n>1$ it provides us with a kind of 
multi-parameter, or ``multi-time'',  
manifestly covariant 
generalization of the Hamiltonian formalism. 
In doing so fields are treated not as 
infinite dimensional mechanical systems 
evolving with time, but rather as systems varying in space-time, 
with the DW Hamiltonian function controlling such a variation 
(similarly to the usual Hamiltonian controlling the time evolution). 

The objective of the present contribution is to discuss an approach 
to quantization of fields based on polymomentum canonical theories. 
Although we confine ourselves exclusively to the approach based on the 
DW theory, we believe that basic ideas presented in what follows 
can be  extended 
to more general polymomentum theories.

\section*{Graded Poisson bracket and quantization}

The canonical quantization in mechanics is essentially 
based on the algebraic structure given by the Poisson bracket. 
One of the reasons why  polymomentum canonical theories 
have not been used as a basis of quantization was   
the lack of an appropriate generalization of the Poisson bracket.  
In \cite{ikanat0} we proposed such a generalization 
within 
the  DW theory. The bracket is defined  on horizontal differential 
forms 
$F=\frac{1}{p!}F\lind{\mu}{p}(z^M)    
dx^{\mu_1}\we ... \we dx^{\mu_p} $
of various degrees $p$ ($0\leq p \leq n$),  which   
play the role of dynamical variables 
(instead of  functions in mechanics or functionals in the 
conventional Hamiltonian formalism in  field theory). 
It leads to graded analogues of the Poisson algebra 
structure\cite{ikanat0,bial96}. 
More specifically, the bracket on differential forms in 
DW theory leads to generalizations of 
the  so-called Gerstenhaber algebra \cite{gerst} 
(a graded analogue of the Poisson 
algebra with the grade of an element of the algebra 
with respect to the bracket 
differing 
by one from its grade 
with respect to the multiplication).    
For the purposes of the present paper it suffices to know 
a small subalgebra of the canonical brackets and 
a 
representation of the field equations in terms 
of the bracket operation\footnote{For the reason of a limited space 
we avoid discussing properties of graded Poisson bracket in 
DW theory in details. In what follows we simply chose facts 
which we need and refer the interested reader for more details to 
\cite{ikanat0,bial96,bial97}. }. 

Using the notation $\om_\mu:= (-1)^{(\mu-1)} 
dx^1\we ... \we \what{dx^\mu}\we ... \we dx^n$ 
the canonical brackets in the (Lie) subalgebra of 
forms of degree $0$ and $(n-1)$ read \cite{ikanat0}  
$$\pbr{p_a^\mu\omega_\mu}{y^b}
= 
\delta^b_a , \quad 
\pbr{p_a^\mu\omega_\mu}{y^b\omega_\nu}
=
\delta^b_a\omega_\nu, \quad 
\pbr{p_a^\mu}{y^b\omega_\nu}
= 
\delta^b_a\delta^\mu_\nu , 
\refstepcounter{equation} 
\eqno {  (\theequation a,b,c) } 
$$ 
with  other  brackets vanishing.  
All brackets in (2) reduce to 
the canonical bracket in mechanics when $n=1$; 
in this sense they are canonical and can be viewed  as 
a starting point of  quantization.

Let us adopt the Dirac correspondence rule that 
Poisson brackets go over to commutators divided by 
$i\hbar$ 
and apply it to the canonical brackets (2). 
Note that this is just an assumption: while this 
principle proved to work well for 
the 
usual Poisson 
bracket its precise form and  applicability to 
graded Poisson bracket in DW theory has to be confirmed. 
By quantizing (2a) we immediately conclude that  
$$
\what{p_a^\mu \om_\mu}=i\hbar \der_a,  
$$ 
where $\der_a$ is a partial derivative with respect to 
the 
field variables. 
The commutator corresponding to (2c) leads to  
a realization of $\what{\om}_\mu$  and $\what{p}{}^\mu_a$  
in terms of Clifford imaginary units, 
or Dirac matrices, 
under the assumption that the 
law of composition of operators is the symmetrized Clifford (=matrix) 
product\cite{qs96,bial97} 
\beq
\hat{p}{}^\nu_a = - i\kappa \ga^\nu\der_a ,
\quad 
\widehat{ \omega}_\nu = - \kappa^{-1} \ga_\nu . 
\eeq 
The quantity 
$\kappa$  of the dimension [{\em length}]$^{n-1}$  
appears here on dimensional grounds. Due to the infinitesimal 
nature of the volume element $\om_\mu$ we expect 
the absolute value of $\kappa$   to be ``very large''. 
Hence its relation to the ultra-violet cutoff scale \cite{qs96} can be 
anticipated 
(see also the last section before Conclusion).

Note that  
the realization of operators in terms of 
Clifford imaginary units  
implies a 
certain 
generalization of the formalism of quantum mechanics. 
Namely, whereas  the conventional quantum mechanics 
is built up on complex numbers    
which are  
essentially the Clifford numbers corresponding to the 
{\em one}-dimensional space-time (= the time dimension in mechanics),  
 the present  approach to  quantization of fields 
viewed as 
multi-parameter  Hamiltonian systems 
(of the De Donder-Weyl type) 
makes use of  the hypercomplex (Clifford) algebra of 
the underlying space-time manifold\cite{bt,hestenes}.

In order to guess the form of 
quantum equations of motions within the present approach it is 
important to know how the field equations are 
represented  in terms of the bracket operation and what is the meaning 
of the bracket with the DW Hamiltonian function. 
In fact, the bracket with $H$ exists only for forms of degree 
higher than $(n-1)$ \cite{ikanat0}. Using $(n-1)$-form 
canonical variables appearing in (2) 
DW Hamiltonian equations (1) can be written in Poisson bracket formulation 
as follows \cite{ikanat0} (cf. \cite{bial97})  
\beq
\bd (y^a\om_\mu)=*\pbr{H}{y^a\om_\mu} = * \der H/\der p^\mu_a, 
\quad 
\bd (p_a^\mu \om_\mu) = *\pbr{H}{p_a^\mu\om_\mu} = - * \der H / \der y^a ,  
\eeq 
where $*$ is the Hodge duality operator acting on horizontal forms,  
and $\bd$ is the total exterior differential 
$
\bd F := \frac{1}{p!}\der_M F\lind{\mu}{p}\der_\mu z^M 
dx^\mu\we dx^{\mu_1} \we ... \we dx^{\mu_p}, 
$ 
with $z^M$ denoting the set of variables $(y^a,p_a^\mu,x^\mu)$. 
For  more general dynamical variables represented by $p$-forms $F$ 
we need a notion of the bracket with an $n$-form $H\om$, where 
$\om:=dx^1\we ... \we dx^n$, 
which allows us to write the equations of motion in 
the symbolic form \cite{ikanat0}   
$$
\bd F = \pbr{H\om}{F} + d^h F,
$$ 
where $d^h$ is the exterior differential with respect to the 
space-time (=horizontal) variables. 
Hence, we 
conclude that the DW Hamiltonian 
``generates'' infinitesimal  space-time variations 
of dynamical variables 
corresponding to  the total exterior differentiation, 
much like the Hamilton 
function in mechanics generates the infinitesimal evolution 
along the time dimension.

Now, an analogue of the 
Schr\"odinger equation 
can be expected to have 
a form 
$ \hat{\i } \hat{d } \Psi \sim \what{H} \Psi $, 
where $\hat{\i }$ and $\hat{d }$ denote appropriate analogues 
of the imaginary unit and the exterior differentiation respectively. 
Keeping in mind the above remark on a hypercomplex generalization 
of quantum mechanics appearing here, an analogy between the 
exterior differential and the Dirac operator 
(in fact, the latter is $d - *^{-1}d*$ \cite{bt}), 
and natural 
requirements 
imposed by the correspondence principle,  
the following generalization of the  Schr\"odinger equation 
can be 
formulated 
\cite{qs96,bial97,firststeps}   
\beq
\label{seqcl}
i \hbar \kappa \gamma^\mu \der_\mu \Psi = \what{H} \Psi, 
\eeq
where $\widehat{H}$ is the operator corresponding to the 
DW Hamiltonian function, the constant $\kappa$ 
of dimension [{\em length}]$^{-(n-1)}$ appears again on dimensional 
grounds, and $\Psi=\Psi(y^a,x^\mu)$ is a wave function over the 
 configuration space of  field and space-time variables. 
In the following section we demonstrate that this 
equation fulfills several aspects of the correspondence principle. 
Note  also that it  reproduces the quantum mechanical Schr\"odinger 
equation at $n=1$. 


Let us construct the DW Hamiltonian operator 
for the system of interacting scalar fields $y^a$ in flat space-time 
given by the Lagrangian density 
\beq
L= \half \der_\mu y^a \der^\mu y_a - V(y). 
\eeq
Then the polymomenta and the DW Hamiltonian function are given by 
\beq 
p^a_\mu= \der_\mu y^a, \quad 
H= \half p^a_\mu p_a^\mu + V(y).  
\eeq
DW Hamiltonian field equations take the form 
\beq
\der_\mu y^a =  p_\mu^a,  
\quad \der_\mu p^\mu_a = - \der V/ \der y^a, 
\eeq 
which is essentially  a first order form of 
a system of coupled Klein-Gordon equations. 

By quantizing the bracket 
\beq
\pbr{ p^\mu_a p^a_\mu}{y^b \omega_\nu}= 2 p^b_\nu   
\eeq
we obtain \cite{bial97}  
$$\widehat{p^\mu_a p^a_\mu}= -\hbar^2 \kappa^2 \lapl ,$$ 
where  $\lapl:= \der_a \der^a$ 
is the Laplacian operator in the  space of field variables. 
Thus the DW Hamiltonian operator of  the system of 
interacting scalar fields takes the form  
\beq
\widehat{H} = -\half \hbar^2 \kappa^2 \lapl + V(y) .
\eeq 

Note that for 
a free scalar field 
$V(y)= (1/2 h^2) m^2 y^2,$ 
so that the DW Hamiltonian operator becomes similar 
to the Hamiltonian operator of the 
harmonic oscillator in the space of field variables.  
Its eigenvalues divided by $\kappa$ 
read 
$m_N=m (N+\half)$. 
Separating variables $\Psi(y,x^\mu)= \Phi(x) f(y)$ 
from (5) we obtain 
$$\what{H} f_N = \kappa m_N f_N, \quad 
i\hbar\ga^\mu\der_\mu\Phi= m_N \Phi .$$
Then for a free scalar field any solution of (5) 
is a linear combination of 
\beq
\Psi_{N,\bk, r}(y,\bx,t)  = u_{N,r}({\bk}) 
f_N(y) e^{\epsilon_r (i\omega_{N,\bk}t - i\bk\cdot\bx)}  ,  
\eeq
where 
$\omega_{N,\bk}:=\sqrt{\bk^2 + m_N^2/\hbar^2}$,  
 $u_{N,r}({\bk})$ is a properly normalized constant spinor, 
$\epsilon_r=+1(-1)$ for positive (negative) energy solutions, and  
$f_N$ are eigenfunctions of the harmonic oscillator in $y$-space. 
As a consequence, any Green  function 
of (5) is given by \cite{firststeps}  
\beq 
K^{} (y',\bx{}',t'; y,\bx{},t) 
=  \sum_{N=0}^{\infty}  \bar{f}_N(y') f_N(y) D_N^{} (\bx{}'-\bx, t'-t),
\eeq
where  $D_N^{}$ denotes a Green function of 
the spinor field of  mass $m_N$. In doing so the type 
of the Green   function $D$ should coincide  with the 
type of the Green function $K$. 
Note that at large space-time 
separations $|x'-x|\gg \hbar/m$ 
the contribution of the term with $N=0$ dominates, so that 
the asymptotic space-time behavior of corresponding Green functions 
is that of a spinor particle with mass $\half m$.    
We hope to present a more detailed analysis elsewhere.


\section*{The Correspondence Principle}
 
In this section we discuss three properties 
of Eq. (5) which make it a proper candidate 
to the  Schr\"odinger equation 
within 
the 
polymomentum quantization. 
All three are in fact different aspects 
of the correspondence principle. 
 
Let us recall  first that the DW canonical theory leads 
to its own field theoretic generalization of the Hamilton-Jacobi 
theory \cite{rund,kastrup}. The corresponding Hamilton-Jacobi equation 
is a partial differential equation 
on $n$ functions 
$S^\mu=S^\mu(y^a,x^\nu)$ 
$$
\der_\mu S^\mu+ H(x^\mu, y^a, p_a^\mu = \der S^\mu / \der y^a )=0.
$$
In a simple example of scalar fields (6)  the 
DW  Hamilton-Jacobi  equation 
reads 
\beq
\der_\mu S^\mu = -\half \der_a S^\mu \der_a S_\mu 
- \half \frac{m^2}{\hbar^2} y^2. 
\eeq
Now, if we substitute 
(a hypercomplex analogue of) 
the quasiclassical ansatz 
\beq 
\Psi = R \, \exp (iS^\mu \ga_\mu / \hka ) \eta, 
\eeq  
where $\eta$ is a constant reference spinor,  
to  
(5) and (10) 
 we obtain 
a set of equations 
which can be transformed to 
the form  \cite{bial97}
\beq
\der_\mu S^\mu = -\half \der_a S^\mu \der_a S_\mu 
- \half \frac{m^2}{\hbar^2} y^2 
+\half \hbar^2 \kappa^2 \frac{\lapl R }{R}, 
\eeq
$$
\der_a S^\mu \der^a S_\mu=\der_a|S| \der^a |S| ,  
\quad    
\der_\mu S^\mu = \frac{S^\mu}{|S|} \der_\mu |S| . 
 \refstepcounter{equation} 
\eqno {  (\theequation a,b) }
$$
In the  first of these we recognize the  
DW Hamilton-Jacobi equation 
(13) with an additional 
term $\half \hbar^2 \kappa^2 \lapl R/{R}$   
which is similar to the 
so-called {\em quantum potential} known in quantum mechanics \cite{bohm} 
and vanishes in the classical limit $\hbar\rightarrow 0$.  
Last two equations are supplementary conditions which  appear 
most likely  due to the fact that the quasiclassical 
ansatz (14) does not represent a most general spinor, thus 
imposing  certain restrictions on  dynamics of the wave function. 
Note that in the 
case of quantum mechanics, $n=1$, conditions (16a,b) reduce 
to trivial identities. 

Thus, it is argued 
that 
in the classical limit equation (5) leads to the DW Hamilton-Jacobi 
equation (with two supplementary conditions 
which are specific to field theory 
and probably are due to restrictions 
imposed by the chosen in (14) analogue of the quasiclassical ansatz).

Another aspect of the correspondence principle 
we are  to consider 
is the Ehrenfest theorem. Let us assume that  
expectation values of operators are given by 
\beq
\left < {} \what{O}{}\right > 
:= \int dy  \, \overline{\Psi}\what{O}{\Psi} , 
\eeq 
where $\overline{\Psi}$ is the Dirac conjugate of $\Psi$. 
These expectation values depend on 
space-time points 
as the averaging is performed only over the field space.   
Using 
generalized Schr\"odinger equation (5) 
with the DW Hamiltonian (10)  
we can show that \cite{bial97}  
\beq 
\der_\mu \left < {}\hat{p}{}_a^\mu {}\right > 
=- \left < {}\der_a \what{H} {}\right > ,  \\ 
\quad \der_\mu\left  <\widehat{y_a \omega^\mu} {}\right > 
= \left  < \what{p^\mu_a \omega_\mu}{}\right > . 
\eeq 
By comparing (18) with  DW Hamiltonian field 
equations (8) we conclude that the latter are fulfilled 
"in average" as a consequence of the representation of operators (3), 
generalized Schr\"odinger equation (5),  
and the definition of 
expectation values (17). 
However, it should be noted that 
this property is fulfilled only for specially 
chosen operators (try e.g. to evaluate $\der_\mu \left < {} y^a {}\right >$ 
to see that this will not yield  the desired result 
$\left < {}\hat{p}{^a_\mu}{}\right >$ for scalar fields).  
Moreover, the scalar product 
$
\int dy  \, \overline{\Psi}{\Psi} 
$ 
implied by definition (17) in general is 
not  positive definite 
and depends on points of the space-time. 
Therefore, it can not be used for a probabilistic 
interpretation. 
These drawbacks 
urge us 
to look for a more 
appropriate  version of the Ehrenfest theorem. 
 
An alternative is suggested by the fact that 
generalized Schr\"odinger  
equation (5) possesses 
a 
positive definite and time independent 
scalar product 
\beq
 \int d\bx \int dy \Psib \beta \Psi  , 
\eeq 
where we introduced the notation 
$\ga^\mu =: (\ga^i,\beta)$ $(i,j = 1,...,n-1)$, thus 
explicitly singling out the time variable $t:=x^n$ and the 
time component of $\ga$-matrices: $\beta:=\ga^t$ ($\beta^2=1$).  
The existence of the satisfactory scalar product 
of this kind
implies  that the probabilistic interpretation of the wave function 
which fulfills 
generalized Schr\"odinger equation (5) is 
possible only if a time dimension is singled out. 
The wave function $\Psi(y^a,\bx,t)$ is  interpreted then 
as a probability amplitude of 
obtaining the field value $y$
in the  
space point $\bx$ 
in the moment of time $t$. 
As a result, the theory becomes very much similar to 
usual quantum mechanics of 
a fictious (spinor) particle in the space of variables $(y^a, \bx)$. 

Now, 
new (global) expectation values of operators 
can be defined by  
\beq  
\left  < \what{O} {}\right > 
:= \int dy \int d\bx \Psib \beta  \what{O} \Psi . 
\eeq 
These expectation values depend only on time. 
Using  definition (20) and 
generalized Schr\"odinger equation (5) written in the 
form 
\beq
i\hbar \der_t \Psi = - i \hbar \al^i \der_i \Psi  +  
\frac{1}{\kappa}\beta \what{H} \Psi , 
\eeq 
where 
$ \al^i := \beta \ga^i$,   
we obtain 
\beq
\der_t \left < {}y^a {}\right > =  \left < {} \what{p}^a_t {}\right >, 
\quad  
\der_t  \left < {} \what{p}_a^t {}\right > 
= - \left < {} \what{\der_i p^i_a} {}\right > 
- \left < {} \der_a \what{H} {}\right > . 
\eeq  
Note that in (22) we identified 
$\what{\der_i p^i_a}$ with 
$-2i\hbar\kappa \ga^i \der_a \der_i $. 
This identification is 
consistent with yet another 
aspect  of the  correspondence principle 
a discussion of which follows.

This  aspect is a relation 
between the classical equations of motion and the 
Heisenberg equations of motion of operators.  
From (21) it follows that the time evolution is 
given by the operator 
\beq
\what{\mbox{$\cal E$}}:= 
- i \hbar \al^i \der_i   + \frac{1}{\kappa}\beta \what{H} . 
\eeq 
Then, proceeding according to 
the standard quantum mechanics 
we obtain 
\beq
\der_t y^a = \frac{i}{\hbar} [\Eop,y^a] = \hat{p}^a_t , 
\quad  
\der_t \what{p}^t_a = \frac{i}{\hbar} [\Eop,\what{p}^t_a] 
= - \what{\der_i p^i_a} - \der_a\what{H} ,     
\eeq  
where we  assumed  as before 
\beq
\what{\der_i p^i_a} =  
-2i\hbar\kappa \ga^i \der_a \der_i . 
\eeq
 Hence, 
as a consequence of generalized 
Schr\"odinger equation (5) and the representation of operators (3),   
the Heisenberg equations of motion 
have the same form as 
classical  DW Hamiltonian equations (1) 
written in the form with a singled out time dimension. 



\section*{Relation to the Schr\"odinger wave functional} 


In this section a possible relationship between the    
Schr\"odinger wave functional in 
quantum field theory\cite{hatfield} 
and our 
wave function is 
examined\footnote{The presentation  
here essentially follows an 
unpublished preprint by the author\cite{firststeps}.}. 
We confine ourselves  to the simplest example of 
a free real scalar field. For the seek of simplicity we 
henceforth put $n=3+1$ and $\hbar=1$. 

The idea is as follows. On the one hand, 
the  Schr\"odinger wave functional 
$\Psi([y(\bx)],t)$ is known to be a probability amplitude of 
the field configuration $y=y(\bx)$ to be observed in the moment 
of time $t$. 
On the other hand, 
our wave function $\Psi(y,\bx,t)$  
can be interpreted as a probability 
amplitude of finding the value $y$ of the field 
in the point $\bx$ in the moment of time $t$. 
Hence, the wave functional could in principle be related to 
a certain composition of single amplitudes given by our 
wave function. 

Let us consider the Schr\"odinger functional 
corresponding to the vacuum state of 
a free scalar field\cite{hatfield} 
\beq   
\Psi_0([y(\bx )], t)= \eta
 \exp \left( 
i E_0 t 
-
\frac{1}{2 {}}  \int \! 
\frac{d \bk}{(2 \pi)^{3}}
\ \omk \, \tilde{y}{}(\bk) \tilde{y}{}(-\bk) 
\right) , 
\eeq
where the Fourier expansion  
$y(\bx) = \int \! \frac{d \bk}{(2 \pi)^{3}}\,y(\bk)
e^{i\bk \bx } $
is used, 
$\eta$ is a normalization factor, 
 $ \omega_\bk := \sqrt{m^2 + \bk^2}$,  
 and 
$E_0$ is the 
vacuum state energy 
\beq
E_0= \lim_{V\rightarrow\infty \atop Q\rightarrow\infty} 
\half  \int_V d\bx  \int_Q \frac{d\bk }{(2\pi)^3}\omega_\bk 
\eeq 
which is divergent if either the ultraviolet cutoff $Q$ of the 
volume of integration in  $\bk$-space 
or the infrared cutoff $V$ of the volume of integration over 
$\bx$-space  go to infinity. The symbol $\lim$ has a formal meaning 
throughout.

By replacing the Fourier integral 
by the Fourier series according to 
the rule $\int \frac{d \bk}{(2 \pi)^{3}} \rightarrow 
\lim_{V\rightarrow\infty}\frac{1}{ V } 
\sum_{[\bk]}$, $[\bk] \in {\ZZ}^3$,  
the Schr\"odinger vacuum state functional can be written in the 
form of an infinite product 
of the harmonic oscillator ground state wave functions  
over all cells in $\bk$-space 
\beq
\Psi_0([y(\bx )], t) = \eta 
\lim_{V\rightarrow\infty} \prod_{[\bk] } 
\exp \, \half \left(  i  \omk t 
-  \frac{1}{ V {} } \ \omk {y}{}^2(|\bk |)  \right) . 
\eeq

Now, let us  consider the ground state ($N=0$) wave functions 
(cf. Eq. (11)) 
of generalized Schr\"odinger equation (5) 
for a free scalar field 
\beq
\Psi_{N=0,\bk} (y_\bx,\bx,t) = u_{N=0}(\bk) 
e^{  i  \omega_{0,\bk} t -i \bk \cdot \bx }
e^{-\frac{m}{2 \kappa} y_{}^2} , 
\eeq
where    
$\omega_{0,\bk}=\sqrt{(\frac{m}{2})^2 + \bk^2}$.   
To simplify a subsequent analysis, 
which is in any case of preliminary character,  
we  ignore in what follows the spinor nature 
of the wave function encoded in  $u_{N=0}(\bk)$. 
 Taking into consideration the probabilistic interpretation 
of solutions (29) 
and assuming that 
there are no correlations 
between the field 
values in space-like separated points,  
the amplitude of funding in the vacuum state 
the whole configuration 
$y=y(\bx)$  
can be represented as an infinite product 
of single amplitudes given by 
the ground state solutions (29) with $y_{}=y(\bx)$ 
over all points $\bx$ of the space.  
In order to ensure the spatial  isotropy and homogeneity 
which are expected for the 
vacuum state    
we also have to take  a product over all possible 
values of wave numbers 
because    each  
separate mode with a wave number $\bk$  violates 
these properties. 
This also 
agrees  with an idea of the vacuum state in 
which all possible $\bk$-states are filled. 
Hence, the following symbolic formula for 
the 
approximate 
composed  vacuum  amplitude 
can be written (up to a normalization)
\beq 
\prod_{\bk } \prod_{\bx  } 
e^{  i  \omega_{0,\bk} t -i \bk \cdot \bx } 
e^{-\frac{m}{2 \kappa {}} y(\bx)^2} .  
\eeq 
This 
expression can be 
assigned a meaning if a certain discretization in both 
$\bx$- and $\bk$-spaces is assumed. 
This discretization can be related to finite values of 
cutoff parameters $V$ and $Q$ which imply  
minimal 
volume elements in $\bk$-space and  in $\bx$-space to be, 
respectively, $(2\pi)^3/V=:\xi^3$ and   $(2\pi)^3/Q=:\lambda^3$.  
Then coordinates in $\bx$- and $\bk$-space are 
integers $[\bx]\in \ZZ^3$ and  $[\bk]\in \ZZ^3$ 
such that  $\bx=[\bx]\lambda$, $\bk=[\bk]\xi$.   
The continuum limit formally corresponds to 
$V\rightarrow\infty$ and $Q\rightarrow\infty$,  
however, an analysis of its existence 
in mathematical sense is beyond the scope of the present 
consideration.   
Using this discretization,  
the obvious identity 
$\prod_{\bk} e^{i\bk\cdot\bx}=1$, 
and the Fourier series expansion 
$y ({\bx}) =\frac{1}{V}\sum_{[\bk]}y_{\bk}e^{i\bk\cdot\bx},$ 
we obtain 
\beqa
&& \prod_{\bk }  
e^{  i  \omega_{0,\bk} t
} 
\prod_{\bx  }
e^{-\frac{m}{2 \kappa {}} y({\bx })^2} 
\nn \\
&=& 
\lim_{V\rightarrow\infty \atop Q\rightarrow\infty} 
\prod_{[\bk] } e^{  i  \omega_{0,\bk} t } 
 \prod_{[\bx]} 
\exp\left( -\frac {m}{2 \kappa {}} \frac{1}{V^2}\sum_{[\bq']}\sum_{[\bq'']} 
y_{\bq'}y_{\bq''}e^{i(\bq'+\bq'')\cdot \bx}   \right) 
\nn \\
&=&
\lim_{V\rightarrow\infty \atop Q\rightarrow\infty} 
\prod_{[\bk] } e^{  i  \omega_{0,\bk} t}   
\exp\left( -\frac {m}{2 \kappa {}} 
\sum_{[\bx]}  
\frac{1}{V^2}\sum_{[\bq']}\sum_{[\bq'']} 
y_{\bq'}y_{\bq''}e^{i(\bq'+\bq'')\cdot \bx}   \right) 
\nn \\
&=&
\lim_{V\rightarrow\infty \atop Q\rightarrow\infty} 
\prod_{[\bk] } e^{  i  \omega_{0,\bk} t  }
 \exp\left( -\frac {m}{2 \kappa {}} 
\frac{QV}{(2\pi)^3}\frac{1}{V^2} 
\sum_{[\bq]} y_{\bq}y_{-\bq}\right) 
\nn \\
&=&
\lim_{V\rightarrow\infty \atop Q\rightarrow\infty} 
\prod_{[\bk] } \exp 
\left( {  i  \omega_{0,\bk} t} 
-\frac {m}{2 \kappa {}V} \frac{Q}{(2\pi)^3}y_{\bk}y_{-\bk} \right) 
,  
\eeqa
where in passing to the fourth  line we have taken into  
account that the number of cells both in 
$\bx$- and $\bk$-space 
is equal to $QV/(2\pi)^3$.

Let us compare the composed amplitude 
(31) with the 
standard vacuum functional in the form 
(28).  
Two additional parameters 
$\kappa$ and  $Q$  
appear in (31):  
$Q$ 
is 
an  (infinitely large)  ultra-violet cutoff of the volume in 
$\bk$-space, while  $\kappa$ 
is essentially the inverse 
of an  infinitesimal (or very small) volume element 
in $\bx$-space (cf. Eq. (3)), 
i.e a kind of fundamental length to the power 3. 
 {}From the physical point of view 
it is  quite natural to relate 
the inverse of the  fundamental 
length to the ultraviolet  cutoff. We thus 
identify  $\kappa=Q/(2\pi)^{3}$ 
obtaining  the composed amplitude 
\beq
\lim_{V\rightarrow\infty}  
\prod_{[\bk] } \exp 
\left( {  i  \omega_{0,\bk} t} 
-\frac {m}{2 V} y_{\bk}y_{-\bk} \right) 
\eeq 
which is similar to (28) 
except that in (32) the proper mass $m$ appears instead 
of the frequency $\omega_{\bk}=\sqrt{m^2 + \bk^2}$ 
and $\omega_{0,\bk}$ replaces $\half \omega_{\bk}$.

It is easy to see that the discrepancy between 
(28) and (32) 
disappears in the 
ultra-local limit $|\bk| \ll m$.   
In this limit the two-point Wightman function 
$\left < {} y(\bx_1) y(\bx_2){}\right >$ 
between space-like separated points 
$\bx_1$ and  $\bx_2$ vanishes,   
so that there are no correlations between 
the field values in these points. 
This is, however, exactly the assumption 
which we made when 
writing the approximate composed amplitude in the form (30). 
Hence, in the ultra-local limit 
the composed amplitude constructed from the 
the ground state wave functions obeying generalized 
Schr\"odinger  equation (5) 
is consistent with  
the Schr\"odinger 
wave functional of the vacuum state (28).  
Unfortunately, an attempt to extend this correspondence 
beyond the ultra-local limit 
leads to  a difficulty of 
writing an  expression for the composed amplitude 
similar to (30) which would account for all relevant 
correlations between the field values in space-like  
separated points. 
 
Note, that another important byproduct of our analysis 
in this section  
is a conclusion  that the constant $\kappa$ which appeared in 
(3) and (5) 
on purely dimensional grounds has to be identified 
with an  ultraviolet cutoff scale quantity.

\section*{Conclusion}

Field theories can be viewed as multi-parameter 
Hamiltonian-like systems in which space-time variables 
appear on equal footing as analogues of 
the 
time parameter in mechanics. 
A quantization 
of such a version of the Hamiltonian formalism leads 
to an 
extension of the formalism of quantum mechanics in which the 
Clifford algebra of underlying space-time manifold 
plays a key role similar to that of complex numbers 
in quantum mechanics.  The latter thus appears as a special 
case of a theory with a single (time) parameter. 
In this formulation a description of 
quantized fields is achieved in terms of  
a (spinor) wave function on a finite dimensional analogue 
of the configuration space 
(the space of field and space-time variables). 
The wave function  satisfies a multi-parameter 
covariant generalization of the Schr\"odinger 
equation, Eq. (5), which is a partial derivative equation 
similar to the Dirac equation with the mass term replaced 
by an operator corresponding to a multi-parameter (polymomentum) 
analogue of Hamilton's canonical function. 
Note that despite the dynamics is formulated in a manifestly 
covariant manner 
the consideration of 
scalar products 
suggests that a proper probabilistic 
interpretation of the wave function 
still may require a time parameter to be singled out.

The  description outlined above appears to be very different from  
that  known in contemporary  quantum field theory.  
A relation to the latter is a challenge  to the theory presented here.  
In this paper we pointed out  a relation 
to the Schr\"odinger wave functional 
which can thus far be followed only in  ultra-local approximation.  
However, the latter is too rough for the real physics.  
Hence, further efforts are required to clarify possible 
connections with the standard quantum field theory.  

Note in conclusion, that 
the present approach may have interesting applications to the 
problem of quantization of gravity and 
field theories on non-Lorentzian  space-times 
if  the problems with the physical interpretation   
and the relationship to the standard 
quantum field theory are  resolved. 
Further discussion  can be found  in \cite{potsdam98} 
were a sketch of an approach to quantization of general relativity 
based on      the present framework is presented.


\end{document}